\documentclass[10pt,journal,compsoc]{IEEEtran}
\usepackage{ctable}

\ifCLASSOPTIONcompsoc
  \usepackage[nocompress]{cite}
\else
  \usepackage{cite}
\fi
\ifCLASSINFOpdf
\else
\fi
\usepackage{pdfpages}
\usepackage{graphicx}
\usepackage{float}
\usepackage{amssymb,amsfonts,amsmath,amsthm}

\usepackage{breqn}
\usepackage{algorithm,algpseudocode}
\usepackage{enumitem}
\usepackage{hyperref}
\usepackage{cleveref}
\begin{document}
%
\title{An Approach for Link Prediction in Directed Complex Networks based on Asymmetric Similarity-Popularity}
%
%
%
%

\author{Hafida~Benhidour, Lama~Almeshkhas, Said~Kerrache
\IEEEcompsocitemizethanks{\IEEEcompsocthanksitem King Saud University, College of Computer and Information Sciences, Riyadh, 11543, KSA.\protect\\
E-mail: hbenhidour@ksu.edu.sa}
}

%
%


\IEEEtitleabstractindextext{%
\begin{abstract}
Complex networks are graphs representing real-life systems that exhibit unique characteristics not found in purely regular or completely random graphs. The study of such systems is vital but challenging due to the complexity of the underlying processes. This task has nevertheless been made easier in recent decades thanks to the availability of large amounts of networked data. Link prediction in complex networks aims to estimate the likelihood that a link between two nodes is missing from the network. Links can be missing due to imperfections in data collection or simply because they are yet to appear. Discovering new relationships between entities in networked data has attracted researchers' attention in various domains such as sociology, computer science, physics, and biology. Most existing research focuses on link prediction in undirected complex networks. However, not all real-life systems can be faithfully represented as undirected networks. This simplifying assumption is often made when using link prediction algorithms but inevitably leads to loss of information about relations among nodes and degradation in prediction performance. This paper introduces a link prediction method designed explicitly for directed networks. It is based on the similarity-popularity paradigm, which has recently proven successful in undirected networks. The presented algorithms handle the asymmetry in node relationships by modeling it as asymmetry in similarity and popularity. Given the observed network topology, the algorithms approximate the hidden similarities as shortest path distances using edge weights that capture and factor out the links' asymmetry and nodes' popularity. The proposed approach is evaluated on real-life networks, and the experimental results demonstrate its effectiveness in predicting missing links across a broad spectrum of networked data types and sizes. 
\end{abstract}

}

\maketitle

\IEEEdisplaynontitleabstractindextext

%
\IEEEpeerreviewmaketitle

\IEEEraisesectionheading{\section{Introduction}\label{sec:introduction}}

Coupled biological and chemical systems, neural networks, the World Wide Web, and the Internet are a few examples of systems composed of highly interconnected units. Those systems are best described using graphs models that are neither purely regular nor completely random which are known as complex networks \cite{albert2002statistical1}. The study of complex networks has attracted the attention of many researchers in various branches of science and engineering, and considerable efforts have been made to understand complex networks' evolution, structure, and function. A critical issue that the research community faces is the incapacity to observe all relationships due either to limitations in the data collection process or because some relationships are to appear shortly in the future \cite{guimera2009missing}. The process of identifying the yet unformed links is called \textit{link prediction}.   

In recent years, the link prediction problem has attracted increasing attention. Link prediction aims to predict the possible links that can form in the future between nodes based on the currently observed links. This problem is crucial both theoretically and practically. In theory, link prediction can help understand the mechanism of complex networks' evolution. Practically, link prediction can be applied in various fields, such as in biological networks, to infer the existence of a link, which will save time and cost. In social networks, link prediction can recommend new friends and enhance the user's experience. Link prediction can also help design efficient recommendation algorithms that include only the relevant information. 

Due to the complexity of link prediction, researchers have proposed many prediction approaches over the years. Not all these algorithms, however, are tailored to complex networks \cite{lu2011link}. The nature and topological characteristics of complex networks, such as hierarchical organization, the existence of popular nodes, and communities can be exploited to improve link prediction performance. 
Furthermore, most previous research focused on the link prediction problem in undirected networks, mainly due to their simplicity. However, not all real-life systems can be represented as undirected networks, and the symmetrization of such networks inevitably results in losing information about the relationships between the system's entities. 

Despite the prevalence of directed networked data such as in communication networks, social networks, trust networks, and biological networks, research on directed link prediction methods remains limited. 
This paper introduces a novel link prediction approach explicitly designed for directed networks. The proposed approach falls into the category of similarity-popularity methods, which have recently been proven successful in undirected networks \cite{Boguna2021, kerrache20}. Unlike in the undirected case, the proposed algorithms handle edge asymmetry by using asymmetric similarity and popularity measures. The algorithms generalize observed similarities to disconnected nodes using shortest path distances. To prove its performance, we experimentally evaluate the proposed method on a broad range of network types and sizes. 

The rest of this paper is organized as follows. Section \ref{sec:problem} gives a formal definition of the link prediction problem. Section \ref{sec:related} contains an overview of existing link prediction approaches, including those specifically designed for directed networks. The proposed method is presented in details in Section \ref{sec:proposed} and evaluated experimentally in Section \ref{sec:experimental}. Finally, Section \ref{sec:conclusion} concludes this paper and presents some future research directions. 

\section{Problem statement}
\label{sec:problem}
Link prediction lies at the intersection of several research fields including machine learning, data mining, recommended systems, and network science \cite{huang2005link, li2009recommendation, liu2007predicting}. This versatility is due to the fact that many tasks related to networked data, such as product and friend recommendation or prediction of research paper citations and protein interactions, can be reduced to a link prediction task. Formally, a link prediction problem is defined by giving a graph $G = (V, E)$, where $V$ is the set of nodes, and $E \subseteq  V \times V$ is the set of edges. We call $E$ the set of observed edges and denote by $ U $ the set of unobserved edges, where $U= V \times V -E$. The goal is to predict which elements of $U$ should belong to $E$. The omission of some edges from $ E $ can be attributed to limitations in the data collection process or because these edges will appear shortly. In the directed network case, the set $ V \times V $, and consequently $ E $ and $ U $, consists of ordered couples so that the edge $(i,j) \neq (j,i)$.

\section{Related Work}
\label{sec:related}
This section will first present link prediction methods for undirected networks. Many of these methods can be used or adapted to directed networks but do not exploit the asymmetric nature of the relationships. We will cover four main link prediction approaches: topological similarity approaches, probabilistic approaches, classification-based approaches, and similarity-popularity approaches. Topological similarity-based methods do not require building a model of the network; they measure the score or distance between each pair of nodes in the network. This score is then used to estimate the possibility that the two corresponding nodes will form an edge.  
Probabilistic methods are usually based on Markov chains and Bayesian networks. They build a model for the whole network, which is then used to estimate the connection probability between disconnected pairs of nodes. 
Classification-based approaches predict links by training a classifier to discriminate between connected and disconnected couples using various topological features.
Similarity-popularity methods assume that the connection between nodes is driven by two factors: popularity, which is an intrinsic property of the node, and similarity between nodes. 

Finally, we will present link prediction methods specifically designed for 
direct networks. 

\subsection{Topological similarity methods} 
Similarity-based methods assume that nodes tend to connect to other similar nodes. They assign a similarity score to non-existing couples with the convention that pairs with a higher score are more likely to appear than those with a lower score. The similarity measure depends on the network's topological properties, but its form varies from one approach to another. Similarity-based methods can be divided according to the amount of information required to compute the score for a given couple of nodes into three categories: local, global and quasi-local methods.

When the score is calculated based on global methods, which consider the whole network topology, the cost to predict all edges can become prohibitive \cite{lu2011link}. On the other hand, a score based on local methods considers only the immediate neighbors of each pair of nodes, which reduces the computational cost and makes them scalable to large complex networks. Lastly, quasi-local methods balance local and global approaches. They typically require more computations than local methods but offer higher prediction accuracy \cite{lu2011link}.

\subsubsection{Local methods}
Local similarity-based methods rely on the nodes' neighborhood structure to calculate the similarity to other nodes in the network. Local methods are generally fast and highly parallelizable, but their main drawback is that they fail to predict the formation of edges between nodes at a distance greater than two. Various local similarity measures have been proposed in the literature, including Adamic-Adar Index \cite{adamic2003friends}, Common Neighbors Index
\cite{newman2001clustering}, Hub Depromoted Index  \cite{ravasz2002hierarchical}, Hub Promoted Index \cite{ravasz2002hierarchical}, Jaccard Index \cite{jaccard1901etude},  Leicht-Holme-Newman Index \cite{newman2001clustering}, Preferential Attachment Index \cite{newman2001clustering}, Salton Index \cite{dillon1983introduction}, and Sorensen Index \cite{sorensen1948method}. 

Authors in \cite{zhou2009predicting}  compared the performance of nine local similarity-based methods on six real networks. The results show that Common Neighbors has the best overall performance, followed by Adamic-Adar. They then proposed a new similarity method inspired by resource allocation, and it gave better results than Common Neighbors while not requiring more information or computational time. This method uses information about the next nearest neighbors and enhances the algorithmic accuracy. Authors in \cite{lu2009similarity}  used a local path index to score node similarity, which strikes a compromise between computational requirements and prediction accuracy. 

\subsubsection{Global methods}
Global similarity-based methods use the entire network topological information to score the edges. The advantage of these methods is that they can predict long-range edges that are impossible to predict with local methods. However, they have two main drawbacks. First, they scale poorly to large networks because of the intense computation required. The second drawback is that their parallelization can be very complex. 
Katz Index (KI) is a global link prediction method that sums the influence of all paths between two nodes, incrementally penalizing paths by their length \cite{katz1953new}. Katz Index and all global methods generally consider paths of length greater than or equal to two. 

The Random Walk method \cite{liu2010link} assigns a couple $(i,j)$ a score defined as the probability that a random walk that starts at $i$ reaches $j$. Random Walk with Restart \cite{tong2006fast} is similar to Random Walk but introduces a probability of returning to the source node and restarting the walk, which results in higher scores for nodes near the source. Average Commute Time \cite{fouss07} is the average number of steps a random walker starting at node $i$ takes to reach node $j$ and return to $i$. Flow Propagation \cite{vanunu2008propagation} is a variation of Random Walk with Restart where the normalized Laplacian matrix replaces the normalized adjacency matrix. SimRank \cite{jeh2002simrank} calculates the similarity between two nodes as the expected time it takes two random walks that start at the two nodes to meet.

\subsubsection{Quasi-local methods}
Quasi-local methods aim at balancing local and global topological similarity measures. They exhibit local methods' computational efficiency and global methods' topological visibility. They are not computationally overwhelmed since they do not consider the similarity between all pair of nodes, but neither are they limited to neighbors of neighbors. The most popular quasi-local methods are based on random walk models \cite{liu2010link} and path counting \cite{lu2009similarity}. In general, quasi-local scores can be seen as generalizations of local scores and can become equivalent to them by limiting specific depth parameters or computation steps. 
The Local Path Index \cite{lu2009similarity} is a quasi-local method strongly based on the Katz index but considers only paths of limited length. Local Random Walk uses the random walk technique but limits the walk to a specific number of iterations \cite{liu2010link}. Superposed Random Walk \cite{liu2010link} emphasizes nodes in the local neighborhood by continuously releasing a new walker from the source node.

\subsection{Probabilistic methods}
Probabilistic approaches assume that the network has a known structure and build a model to fit this structure by estimating its parameters using statistical methods. The model is then used to estimate the probability of non-existing edges \cite{goldenberg2010survey}. 

The Hierarchical Random Graph model  \cite{clauset2008hierarchical}  is a probabilistic model that exploits the fact that many real networks exhibit a hierarchical structure. In these kinds of networks, nodes with a high degree have a lower clustering coefficient than low-degree nodes. High-degree nodes weakly connect highly clustered nodes within isolated communities, thus forming a hierarchical structure. Clauset et al. \cite{clauset2008hierarchical} proposed a method to represent a hierarchical structure using a dendrogram where each leaf represents a node in the network, and each internal node represents a cluster. Each internal node of the dendrogram is associated with the probability that a link exists between any of its children. The probability that two nodes are connected can then be computed by finding their lowest common ancestor.  

Some networks in the real world do not fit in the hierarchical schema. The Stochastic Block Model assumes instead that the nodes are distributed in communities or blocks. The model restricts the communities from overlapping, and the probability of link formation between any two nodes depends on the block to which they belong \cite{guimera2009missing}.

The Fast Blocking Probabilistic Model \cite{liu2013correlations} is based on partitioning the network into a set of blocks of either a community or a particular community composed of small degree nodes. The model uses a greedy strategy to partition the network into communities efficiently. The connection probability between a couple of nodes is estimated by computing the link density between and within communities over a set of partitions. This implies that two nodes will have a higher chance of forming a link if they belong to the same community. 

The Cycle Formation Model is based on the assumption that the network aims to have close cycles in its formation process \cite{huang2010link}. Such assumption matches the Common Neighbors method, which counts the number of cycles of length three that could be formed. This method captures the longer cycles by extending the clustering coefficient to a generalized clustering coefficient of higher order. This method can thus be seen as a generalization of the Common Neighbors method to longer cycles.

\subsection{Classification-based approaches}
Link prediction can be seen as a classification problem with two classes: one is positive, and the other is negative. The positive class represents connected pairs of nodes, whereas the negative class consists of disconnected couples. Hasan et al. \cite{al2006link} used local features such as the short distance between two nodes, aggregated features such as a sum of neighbors, and semantic features such as the number of matching keywords to discriminate between disconnected and connected couples. They used these features to build a classifier model that helps predict author collaboration and evaluated their approach on two co-authorship networks. They assessed each feature visually by comparing the density of the classes and then using well-defined feature ranking methods. As a result, they found that semantic similarity helps improve link prediction accuracy.     

Link prediction is arguably an imbalanced problem where the two classes exhibit vast differences in size. Such extreme class skew reduces the performance of link prediction algorithms. There are generally two approaches to deal with this issue: downsampling the majority classes or oversampling the minority class. However, this approach might cause data loss in downsampling and noise in data oversampling. The second solution is reducing the problem to a cost-sensitive learning problem with unknown miss-classification costs as proposed in \cite{doppa2010learning}. The authors tackled this issue by proposing a new cost-sensitive formulation based on chance constraints. They tuned the cost to find the hypothesis that best separates the positive and negative instances and reduces the miss-classification cost. They performed a detailed evaluation of their approach using networks from three different domains. They found that they indeed can effectively formulate the link production as a complex structured prediction problem with many constraints.

\subsection{Similarity-popularity methods}
Similarity-popularity methods ascribe network topology to the similarity between nodes and their popularity, which is their capacity to connect to other nodes. Their assumption is that the more similar and popular the two nodes are, the higher their connection probability.

The complex network model proposed in \cite{boguna2009navigability} assumes the existence of a hidden metric space that underlies the network and governs the similarity between nodes. We refer to this model as the Hidden  Metric  Space Model (HMSM). The probability of connecting two nodes in an undirected network based on HMSM is defined as follows:
\begin{equation} 
\label{eq:hmm_eq}
r_{ij} = \left(1+\frac{d_{ij}} { \phi (\kappa_i, \kappa_j)}\right)^{-\alpha} ,
\end{equation}
where $ d_{ij} $  is the distance between nodes $ i $ and $ j $, $ \kappa_i $ and $ \kappa_j $ denote their expected degrees, and the characteristic distance scale $ \phi (\kappa_i, \kappa_j) \approx \kappa_i \kappa_j $. The parameter $ \alpha $ represents the influence of the hidden metric space on the observed topology. Large values of $ \alpha  $ lead to a strong clustering in the network. The connection probability $ r_{ij} $ increases when the distance $ d_{ij} $ becomes small; similar nodes (characterized by a small distance in the hidden space) have a high probability to connect. The probability $ r_{ij} $ also increases when $ \kappa_i  \kappa_j $ becomes large. This is reflected in real  networks where popular nodes (characterized by high degrees) tend to connect to many other nodes.

The authors of \cite{papadopoulos2012popularity} explored the trade-off between popularity and similarity in growing networks. Nodes in growing networks prefer to link to the most popular nodes and the closest ones to them. The authors compared their results with real-world complex networks, showing that the probability of forming new links was easily predicted with high accuracy. 
Following \cite{papadopoulos2012popularity}, the authors in \cite{zou2014exploiting} worked on two approaches to predict links in the Twitter network. The first is by combining rankings generated by popularity and similarity methods. The second approach is adapting collaborative filtering methods to combine popularity and similarity. The results showed that popularity-based methods generally achieve better performance. Authors in \cite{fastlp} proposed FastLP, a heuristic method that uses geodesic distance to estimate the distance between the nodes in the hidden metric space. It also uses the hidden metric space model \cite{boguna2009navigability} to estimate the nodes linking probabilities. The authors suggested multiplying the geodesic distance with a factor that gives the paths with more edges higher weights. The results show that their FastLP approach can handle large networks without reducing the prediction performance.

In \cite{kerrache20}, the authors propose a link prediction method where in addition to similarity and popularity, link formation also depends on attraction induced by the local neighborhood. In this approach, node popularity and local attraction are directly estimated from the observed network topology and factored out using a specific edge length map. The latter is then used to approximate the distance between non-adjacent nodes via shortest path distances.The method delivers highly accurate predictions at a low computational cost compared to global methods.

\subsection{Link prediction in directed networks}
Most research on link prediction focuses on undirected networks, mostly to their simplicity compared to directed networks. Nonetheless, some authors tackled the problem explicitly in the directed case \cite{schall2014link,garcia2014link}. 

In \cite{schall2014link}, the authors proposed a link prediction technique using a graph pattern called Triadic Closeness to predict links between nodes in directed graphs. The approach assigns a probability to edges that might be closed in the future based on node weight. The weighting procedure distinguishes between the popular neighbors and the neighbors unique only to a few users by giving the latter a higher weight. The authors also designed a reusable framework to handle different social networks. The framework consists of three layers:  data, prediction, and presentation. The data layer is concerned with low-level data handling and persistence management. The prediction layer groups the logic of metric calculations and triad detection. The similarity algorithms and evaluation metrics are applied in this layer to compare the results. Finally, the presentation layer provides visualization and export capabilities. The results show that the pattern-based approach gives better results than local algorithms. 

Authors in \cite{garcia2014link} tacked the problem of link prediction in large networks where algorithms suffer from low scalability and precision problems. As a trade-off between these two issues, the authors proposed a link prediction algorithm that exploits the hierarchical properties of directed networks. The proposed method can be classified as a local method for scoring the edges, making it scalable to large networks. They assume the existence of a local hierarchy which results in better prediction precision. They demonstrated the effectiveness and efficiency of their method on two hierarchical datasets and one non-hierarchical dataset.

Compared to research on link prediction in undirected networks, link prediction in directed networks has received little attention from the research community. There is a clear need for methods specific to directed networked data that can handle and exploit the asymmetric nature of node relationships to produce more accurate predictions, which constitutes the aim of this work.

\section{The proposed algorithms}
\label{sec:proposed}
The hidden metric space model \cite{boguna2009navigability, Serrano2008Self-similarityofComplexNetworksandHiddenMetricSpaces, Boguna2021, Ortiz2017} is a similarity-popularity model initially proposed for undirected networks. It assumes that nodes tend to connect if they are popular or similar; that is, they are close to each other in the hidden metric space. This idea inspires the first link prediction algorithm we propose in this paper. We assume that each node $i$ has a hidden in-degree $\kappa_{i}^{in}$  and a hidden out-degree $\kappa_{i}^{out}$ in addition to the hidden distance. The probability that a node $i$ connects to a node $j$ is then given by
\begin{equation} 
\label{eq:hmm_direc_eq}
r_{ij} = \left( 1 + \frac{d_{ij}}{ \phi(\kappa_{i}^{out}, \kappa_{j}^{in})}\right)^{-1},
\end{equation}
where $ d_{ij}$ is the hidden distance between nodes $i$ and $j$, and $\phi$ is the characteristic distance scale function that should be increasing in its two parameters. We propose to use the following characteristic distance scale function:
\begin{equation} 
\phi(\kappa_{i}^{out}, \kappa_{j}^{in}) =  \dfrac{\log({\kappa}_i^{out}+1) + \log({\kappa}_j^{in}+1)}{\log({\kappa}_{max}^{out}+1) + \log({\kappa}_{max}^{in}+1)},
\end{equation}
where ${\kappa}_{max}^{out}$ and ${\kappa}_{max}^{in}$ are the maximum in-degree and out-degree in the network respectively. 
Hence, the probability of the directed edge $(i,j)$ increases when $i$ has a large hidden out-degree and $j$ has a large hidden in-degree.
Using this formula implies that, in general, $r_{ij} \neq r_{ji}$ even if the distances are symmetric.
Note that, in contrast to the original hidden metric space model, $\alpha$ is set to 1 since we are not interested in controlling the clustering coefficient of the network. 

Since the metric space is hidden, we will use the partially observed topology to estimate the hidden distances $d_{ij}$. To achieve this, we assume that the hidden degree of the node equals the observed one, that is, $ \kappa_{i}^{out} = \hat{\kappa}_{i}^{out} $ and $ \kappa_{i}^{in} = \hat{\kappa}_{i}^{in} $,  where $ \hat{\kappa}_{i}^{out} $ and $ \hat{\kappa}_{i}^{in} $ are the observed out-degree and in-degree of node $ i $ respectively. 
For each edge $ (i,j) $ in the network, we assign the weight $w_{ij}$  defined as follows:
\begin{equation} 
\label{eq:alg1-w}
w_{ij} =  \phi(\hat{\kappa}_{i}^{out},\hat{\kappa}_{j}^{in}).
\end{equation}
Hence edges between popular nodes are assigned larger weights than edges linking unpopular nodes. The rationale is that the connection between two nodes with a large degree can be explained by their popularity and does not require high similarity between them. Distance between non-adjacent nodes is computed as the (directed) shortest path distance between them. If no path exists between two nodes, their distance is set to $+\infty $. 
We use Dijkstra's algorithm to calculate the shortest path distances due to its low computational time complexity compared to other algorithms. To reduce the computational overhead further, we follow \cite{kerrache20} and impose a horizon depth limit on the shortest paths. Namely, we run Dijkstra's algorithm up to $h$ edges from the source node; Any nodes that cannot be reached by up to $h$ edges are assigned infinite distance. Finally, the directed edge $(i,j)$ is assigned the score:
\begin{equation}
s^{ALG1}_{i j} = \left( 1 + \dfrac{d_{ij}}{\phi(\hat{\kappa}_{i}^{out},\hat{\kappa}_{j}^{in})}\right)^{-1}
\end{equation}
The resulting link prediction algorithm is shown in Algorithm \ref{alg:1}. 
\begin{algorithm}[!t]
	\caption{}
	\label{alg:1}
	\begin{algorithmic}[1]
		\For{each node $i \in$ V}		
		\State $\hat{\kappa}_{i}^{in} \gets$ the observed in-degree of node $ i $,
		\State $\hat {\kappa}_{i}^{out} \gets$ the observed out-degree of node $ i $,
		\EndFor
		\For{each link $(i,j) \in E$}
		\State $w_{ij} \gets  \dfrac{\log(\hat{\kappa}^{out}_i +1)+\log(\hat{\kappa}^{in}_j +1)}{\log(\hat{\kappa}^{out}_{\max} +1) + \log(\hat{\kappa}^{in}_{\max} +1)}$,
		\EndFor
		\State $ \{d_{ij}\}\leftarrow$\textbf{ShortestPathDistance}($V, E, \{w_{ij}\},h$)
		\For{each couple $(i,j) \notin E$}
		\State $ \pi_{ij} \leftarrow \dfrac{\log(\hat{\kappa}^{out}_i +1)+\log(\hat{\kappa}^{in}_j +1)}{\log(\hat{\kappa}^{out}_{\max} +1) + \log(\hat{\kappa}^{in}_{\max} +1)}$,
		\State $s^{ALG1}_{ij} \gets \left( 1 + \dfrac{d_{ij}}{\pi_{ij}}\right)^{-1}$,
		\EndFor
		\State \Return  $\{s^{ALG1}_{ij}\}$
	\end{algorithmic}
\end{algorithm}

The second algorithm we propose in this paper is an adaptation of the one proposed in \cite{kerrache20} to directed networks. This approach is also a similarity-popularity link prediction method but assumes an additional factor that affects the formation of links: local attraction. More precisely, the likelihood of a link joining two nodes $i$ and $j$ is assumed proportional to
\begin{equation}
\label{eq:alg2}
s^{ALG2}_{i j} = \left(\pi_{ij} + \eta_{ij}\right) \sigma_{ij},
\end{equation}
where $\sigma_{ij}$ is the similarity between $i$ and $j$, $ \pi_{ij} $ is the popularity term, and $ \eta_{ij} $ represents the attraction induced by the local neighborhood. Following \cite{kerrache20}, we define the similarity between the two nodes $i$ and $j$ as
\begin{equation}
\sigma_{ij} = \dfrac{1}{1+d_{ij}},
\end{equation}
To adapt the method to directed networks, we set the popularity term to:
\begin{equation} 
\pi_{ij} =  \dfrac{\log({\kappa}_i^{out}+1) + \log({\kappa}_j^{in}+1)}{\log({\kappa}_{max}^{out}+1) + \log({\kappa}_{max}^{in}+1)}.
\end{equation}
Finally, we define the local attraction term as:
\begin{equation}
\label{eq:eta}
\eta_{ij} \leftarrow 1 - \dfrac{\eta^{in}_{ij}\eta^{out}_{ij}+\eta^{in}_{ij}+\eta^{out}_{ij}}{3},
\end{equation}
where:
\begin{align}
\label{eq:eta-in-out}
	\eta^{in}_{ij} &= \prod_{k \in \Gamma_{ j}^{out} \cap \Gamma_{ i}^{in}} \dfrac{\log(\hat{\kappa}_{k}+2)}{\log(\hat{\kappa}_{\max}+2)},\\
	\eta^{out}_{ij} &= \prod_{k \in \Gamma_{i}^{out} \cap \Gamma_{ j}^{in}} \dfrac{\log(\hat{\kappa}_{k}+2)}{\log(\hat{\kappa}_{\max}+2)}.
\end{align}

Note that using this definition, we focus on the nodes that lie on a directed path of length two, starting at $i$ and ending at $j$ and vice versa. The lower the degree of these nodes, the more strongly $i$ and $j$ are attracted.

Similar to Algorithm \ref{alg:1}, we introduce a weight map on the network, were each edge link $(i,j) \in E$ is assigned the weight:
\begin{equation}
\label{eq:alg2-w}
	w_{ij} = \dfrac{2\pi_{ij}}{1+ \eta_{ij}}.
\end{equation}
Note that due to the asymmetry in the definition of $\pi_{ij}$ and $\eta_{ij}$, we have that $w_{ij} \neq w_{ji}$ in general. This weight definition assigns larger weights to nodes with low local attraction. The rationale is that nodes with high local attraction are likelier to belong to the same neighborhood and hence have high similarity. The resulting method is shown in Algorithm \ref{alg:2}.

\begin{algorithm}[!t]
	\caption{}
	\label{alg:2}
	\begin{algorithmic}[1]
		\For{each node $i \in$ V}		
			\State $\hat{\kappa}_{i}^{in} \gets$ the observed in-degree of node $ i $,
			\State $\hat{\kappa}_{i}^{out} \gets$ the observed out-degree of node $ i $,
			\State $\hat{\kappa}_{i} \gets \hat{\kappa}_{i}^{in}+ \hat{\kappa}_{i}^{out}$,
		\EndFor
		\For{each link $(i,j) \in E$}
			\State $ \pi_{ij} \leftarrow \dfrac{\log(\hat{\kappa}^{out}_i +1)+\log(\hat{\kappa}^{in}_j +1)}{\log(\hat{\kappa}^{out}_{\max} +1) + \log(\hat{\kappa}^{in}_{\max} +1)}$,
			\State $\eta^{in}_{ij} \leftarrow 1 - \prod_{k \in \Gamma_{ j}^{out} \cap \Gamma_{ i}^{in}} \dfrac{\log(\hat{\kappa}_{k}+2)}{\log(\hat{\kappa}_{\max}+2)}$,
			\State $\eta^{out}_{ij} \leftarrow 1 - \prod_{k \in \Gamma_{i}^{out} \cap \Gamma_{ j}^{in}} \dfrac{\log(\hat{\kappa}_{k}+2)}{\log(\hat{\kappa}_{\max}+2)}$,
			\State $\eta_{ij} \leftarrow 1 - \left(\eta^{in}_{ij}\eta^{out}_{ij}+\eta^{in}_{ij}+\eta^{out}_{ij}\right)/3$,
			\State $w_{ij} \leftarrow \dfrac{2\pi_{ij}}{1+ \eta_{ij}}$,
		\EndFor
		\State $ \{d_{ij}\}\leftarrow$\textbf{ShortestPathDistance}($V, E, \{w_{ij}\}, h$)
		\For{each couple $(i,j) \notin E$}
			\State $ \pi_{ij} \leftarrow \dfrac{\log(\hat{\kappa}^{out}_i +1)+\log(\hat{\kappa}^{in}_j +1)}{\log(\hat{\kappa}^{out}_{\max} +1) + \log(\hat{\kappa}^{in}_{\max} +1)}$,
			\State $\eta^{in}_{ij} \leftarrow 1 - \prod_{k \in \Gamma_{ j}^{out} \cap \Gamma_{ i}^{in}} \dfrac{\log(\hat{\kappa}_{k}+2)}{\log(\hat{\kappa}_{\max}+2)}$,
			\State $\eta^{out}_{ij} \leftarrow 1 - \prod_{k \in \Gamma_{i}^{out} \cap \Gamma_{ j}^{in}} \dfrac{\log(\hat{\kappa}_{k}+2)}{\log(\hat{\kappa}_{\max}+2)}$,
			\State $\eta_{ij} \leftarrow 1 - \left(\eta^{in}_{ij}\eta^{out}_{ij}+\eta^{in}_{ij}+\eta^{out}_{ij}\right)/3$,
			\State $ \sigma_{ij} \leftarrow \dfrac{1}{1+d_{ij}}$,
			\State $ s^{ALG2}_{ij} \leftarrow \left(\pi_{ij}+ \eta_{ij}\right) \sigma_{ij}$,
		\EndFor		
		\State \Return  $\{ s^{ALG2}_{ij}\}$	
	\end{algorithmic}
\end{algorithm}

\section{Experimental evaluation}
\label{sec:experimental}
To evaluate the proposed algorithms' performance, we perform several experiments on real network data. We start by assessing the effect of the horizon depth on the performance of our algorithms. We then compare their performance to topological similarity-based link prediction algorithms. In the remainder of this section, we will present the datasets and performance measures used in the evaluation procedure, followed by a presentation of the competing methods and finally a discussion of the obtained results.

\subsection{Datasets}
We use 23 real-life networks of various types and sizes for the experimental evaluation of the proposed algorithms. These networks are available through different public data repositories \cite{snapnets,  konect, pajaknets}. We assess the performance of the methods over a wide range of dataset types, including social, informational, linguistic,  communicational, and biological networks. All used networks are directed with sizes ranging from a few dozens of nodes to over 10,000. Table \ref{tab:data} contains the description of these networks and their number of nodes and links.

\begin{table*}[!t]
	\centering 
	\caption{Description of the networks used in the experimental performance analysis. Columns $n$ and $m$ represent the number of nodes and links in the network, respectively.}
	\label{tab:data}
	
	\begin{tabular}{l p{11cm} r  r } 
		\toprule
		Network & Description & $n$ & $m$ \\
		\midrule
		Adolescent Health\cite{adolescent-health} & Directed friendship network among students. The dataset is available at \url{http://konect.cc/networks/moreno_health}. &2,539 & 12,969\\ \midrule
		
		Advogato\cite{advogato} & The Advogato trust network, where nodes represent users and edges represent trust relationships. Network available at \url{http://konect.uni-koblenz.de/networks/advogato}. &5,155 &39,285\\ \midrule
		
		BitcoinAlpha\cite{bitcoinalpha} & Trust network from the Bitcoin Alpha platform, on which Bitcoins are traded. Only links with positive trust are kept. Network available at \url{http://konect.cc/networks/soc-sign-bitcoinalpha}. &3,683 & 22,650\\ \midrule
		
		Centrality Literature\cite{centrality-90} &  Network centrality citation network from 1948 to 1979. The network is available at \url{http://vlado.fmf.uni-lj.si/pub/networks/data/GD/a01.zip}. &118 & 613 \\ \midrule
		
		Chesapeake Middle \cite{chesapeake2002} &  Middle Chesapeake Bay food web in summer. This network is available at \url{http://vlado.fmf.uni-lj.si/pub/networks/data/bio/foodweb/ChesMiddle.paj}. &37 &198 \\ \midrule
		
		Ciao\cite{ciao} & Trust network from the site \url{http://dvd.ciao.co.uk} during 2013. Network available at \url{http://konect.cc/networks/librec-ciaodvd-trust}. &4,658 & 40,133\\ \midrule
		
		Codeminer \cite{Heymannm2008Javacode} & Th call-graph of a Java program. Nodes represent packages, classes, fields and methods. Edges represent calls. The network is available at \url{https://github.com/gephi/gephi.github.io/tree/master/datasets}. &724 &1,015 \\ \midrule
		
		Criminal\cite{criminal} & Phone calls network between the members of a drug trafficking group. The network is available at \url{https://sites.google.com/site/ucinetsoftware/datasets/mainaseuropoldatasets}. &2,749 &2,952 \\ \midrule
		
		DNA Citation \cite{Hummon1989ConnectivityInCitationNetworkDNA} & DNA research literature citation network. Nodes represent research papers and edges represent citations. This network is available at \url{http://vlado.fmf.uni-lj.si/pub/networks/Data/cite/default.htm}. &39 &61\\ \midrule
		
		DNC Email \cite{konect} &  Network of leaked emails from the 2016 Democratic National Committee. This network is available at \url{http://konect.cc/networks/dnc-temporalGraph}. &1866 & 5517\\ \midrule
		
		FilmTrust \cite{filmtrust} & Trust network from the FilmTrust project. Nodes represent users, and edges represent trust relationships. The dataset is available at \url{http://konect.cc/networks/librec-filmtrust-trust}. &874 &1853 \\ \midrule
		
		FOLDOC\cite{foldoc} & Network of cross-references in the Free On-line Dictionary of Computing (FOLDOC, www.foldoc.org). Nodes represent terms and a directed edge between two terms indicates that the second term is used in the definition of the first one. Network available at \url{http://konect.cc/networks/foldoc}. &13,356 & 120,238\\ \midrule
		
		Human Protein\cite{human-protein} & Network of protein interaction in Humans. The dataset is available at \url{http://konect.cc/networks/maayan-Stelzl}. &1,702 & 6,171 \\ \midrule
		
		Indochina 2004\cite{bovwfi, brsllp} & A web network. The data is available at \url{http://networkrepository.com/web_indochina_2004.php}. &11,358 &47,606\\ \midrule
		
		Japan Air \cite{guimera2005}& Japan air transportation network extracted from the  World Transport network \cite{guimera2005}. This network available at \url{http://seeslab.info/media/filer_public/63/97/63979ddc-a625-42f9-9d3d-8fdb4d6ce0b0/airports.zip}. &56 &183 \\ \midrule
		
		Manufacturing e-mail\cite{michalski2011matching} & Email communication network in a manufacturing company. The network is available at \url{https://www.ii.pwr.edu.pl/~michalski/datasets/manufacturing.tar.gz}. &167 &3,250\\ \midrule
		
		Mathoverflow C2A\cite{paranjape17} & Comment-to-answer interaction network between users on Mathoverflow. Dataset available at \url{https://snap.stanford.edu/data/sx-mathoverflow.html}. &13,778 &71,234\\ \midrule
		
		ODLIS\cite{Reitz2002} & Network of cross-references in the Online Dictionary of Library and Information Science (ODLIS). Nodes represent terms and a directed edge between two terms indicates that the second term is used in the definition of the first one. The dataset is available at \url{http://vlado.fmf.uni-lj.si/pub/networks/data/dic/odlis/Odlis.htm}. &2,900 &16,377 \\ \midrule
		
		Residence Hall\cite{residencehall98} & Directed friendship network between residents living in a residence hall at the Australian National University campus. The network is available at \url{http://moreno.ss.uci.edu/data.html#oz}. &217 &1,839 \\ \midrule
		
		US Air 97 & North American Transportation Atlas Data (NORTAD). The network is available at \url{http://vlado.fmf.uni-lj.si/pub/networks/data/map/USAir97.net}. &332 &2,126 \\ \midrule
		
		Web Edu\cite{gleich04} & A web network.The data is available at \url{http://networkrepository.com/web-edu.php}. &3,031 &6,474 \\ \midrule
		
		Web EPA\cite{pajek11} & Network of web pages linking to \url{www.epa.gov}. Network available at \url{http://networkrepository.com/web-EPA.php}. &4,271 &8,909\\ \midrule
		
		WikiTalk\cite{wikitalk} & Communication network on the Welsh Wikipedia. Nodes are users and an edge indicate that the first user wrote a message on the talk page of the second user. The dataset is available at \url{http://konect.cc/networks/wiki_talk_cy}. &2,101 & 3,951\\ 
		
		\bottomrule
	\end{tabular}
\end{table*}

\subsection{Performance measures}

We will use the three performance measures commonly encountered in the evaluation of link prediction algorithms: the area under the receiver operating curve (AUROC), the area under the precision-recall curve (AUPR), and top-precision (TPR). The AUROC,  which can be computed as the probability that a false negative link is assigned a higher score than a true negative link, has long been the primary performance metric in link prediction literature \cite{Lu2011LinkPredictioninComplexNetworks}. However, due to the severe imbalance in link prediction datasets, especially in large spare networks, authors have recently realized its limitations in evaluating link prediction algorithms. Indeed, the AUROC tends to produce overly-optimistic results due to the severe difference in magnitude between the negative and positive set sizes \cite{yang2015EvaluatingLinkPredictionMethods,garcia2016limitations, Wang2016,muscoloni2017LocalRing, kerrache20}. The AUPR remedies this by focusing only on the positive set, which is particularly useful in link prediction problems \cite{Davis2006TheRelationshipBetweenPrecision-RecallandROCCurves}\cite{yang2015EvaluatingLinkPredictionMethods}.    

TPR, also known as top-$k$ precision \cite{yang2015EvaluatingLinkPredictionMethods}, is computed by first sorting the negative links in decreasing order of the score and then taking the ratio of removed links among the top $k$ ranked links. It is custom to choose $k$ as the total number of removed edges, which means that a perfect link prediction algorithm will have a TPR equal to 1 \cite{Lu2011LinkPredictioninComplexNetworks,yang2015EvaluatingLinkPredictionMethods}. Similar to the AUPR, TPR avoids the pitfalls of the AUROC and has consequently been adopted as the primary performance metric for link prediction algorithms in recent years \cite{Wang2016, muscoloni2017LocalRing, kerrache20}. 
We will follow this practice in this paper and adopt TPR as the primary performance criterion, especially in large networks where its computation is much more efficient than the AUROC and the AUPR. Furthermore, we will use the average significant ranking procedure recently introduced in \cite{kerrache20} to aggregate the performance results over multiple heterogeneous datasets. In this approach, we perform a two-tailed paired t-test to compare the mean performance results of each couple of algorithms on every dataset. If the statistical test reveals insignificance at the specified confidence level, the two algorithms are assigned the score of 0 for the corresponding network; otherwise, the algorithm with the higher performance is assigned the score of 1, and the other algorithm the score of -1. We then rank the algorithms based on the scores obtained for all networks, and in the case of a tie, we use the average rank. An algorithm's final aggregate performance measure is its average rank over all networks, with the convention that the lower the rank, the better. Because of the use of statistical significance tests, this procedure produces a more statistically robust comparison than simply ranking the algorithms as in \cite{Wang2016, muscoloni2017LocalRing}.

\subsection{Competing methods}
\label{sec:directed}
To assess the performance of the proposed algorithms, we compare them to the most widely used similarity methods. However, to carry a fair comparison, we adapt these methods to the directed network settings.
In what follows, $\Gamma^{in}_i$ stands for the set of in-neighbors of $i$ formally defined as $\Gamma^{in}_i =\{k \in V| (k,i) \in E\}$; $\Gamma^{out}_i$ stands for the set of out-neighbors of $i$, that is, $\Gamma^{out}_i =\{k \in V| (i, k) \in E\}$; $\Gamma_i = \Gamma^{in}_i \cup \Gamma^{out}_i$; $\left| \cdot \right|$ stands for set cardinality; $\kappa^{in}_i$ denotes the in-degree of node $i$: $\kappa^{in}_i= \left| \Gamma^{in}_i \right| $; $\kappa^{out}_i$ denotes the out-degree of node $i$: $\kappa^{out}_i= \left| \Gamma^{out}_i \right| $; Finally, $\kappa_i$ denotes the degree of node $i$: $\kappa_i= \left| \Gamma_i \right| $.
\begin{itemize}
	\item Directed Adamic-Adar Index (DADA): Adamic-Adar Index assigns more significant weight to neighbors with fewer neighbors. To handle the case of a directed network, we consider neighbors on a directed path of length two joining the two nodes of interest:
	\begin{equation} 
	\label{eq:dada}
	s^{DADA}_{ij} = \sum_{k \in \Gamma_i^{out} \cap \Gamma_{j}^{in}} {\frac {1}{\log \kappa_k}},
	\end{equation}
	
	\item Directed Common Neighbors (DCNE): Common Neighbors simply counts the number of neighbors common to both nodes. In the directed case, we count only nodes lying on a directed path of length two, starting at the first node and ending at the second:
	\begin{equation} 
	\label{eq:dcne}
	s^{DCNE}_{ij} = |\Gamma_{i}^{out} \cap \Gamma_{j}^{in}|.
	\end{equation}
	
	\item Directed Hub Depromoted Index (DHDI): Given a couple $(i,j)$, the Hub Depromoted Index scales the number of common neighbors (Common Neighbors score) by the maximum between the degree of $i$ and that of $j$, which discourages connections to nodes with high degrees, also known as hubs. In the directed case, we scale the DCNE score by the maximum between the out-degree of $i$ and the in-degree of $j$:
	\begin{equation} 
	\label{eq:dhdi}
	s^{DHDI}_{ij} = \frac {|\Gamma_{i}^{out} \cap \Gamma_{j}^{in}|}{\max (\kappa_{i}^{out} ,\kappa_{j}^{in})}.
	\end{equation}
	
	\item Directed Hub Promoted Index (DHPI): Given a couple $(i,j)$, the Hub Promoted Index scales the number of common neighbors by the minimum between the degree of $i$ and that of $j$,  which to the converse of the Hub Promoted Index encourages connections to hubs. In the directed case, and similar to DHDI, we scale the DCNE score by the minimum between the out-degree of $i$ and the in-degree of $j$:
	\begin{equation} 
	\label{eq:dhpi}
	s^{DHPI}_{ij} = \frac {|\Gamma_{i}^{out} \cap \Gamma_{j}^{in}|}{\min (\kappa_{i}^{out} ,\kappa_{j}^{in})}.
	\end{equation}
	
	\item Directed Jaccard Index (DJID): Jaccard Index assigns higher scores to pairs of nodes that share a higher proportion of common neighbors relative to the number of non-shared neighbors. For a couple (i,j), the directed version we propose focuses on the out-neighbors of $i$ and in-neighbors of $j$: 
	\begin{equation} 
	\label{eq:djid}
	s^{DJID}_{ij} = \dfrac {|\Gamma_{i}^{out} \cap \Gamma_{j}^{in}|}{\kappa_{i}^{out} + \kappa_{j}^{in} -|\Gamma_{i}^{out} \cap \Gamma_{j}^{in}|}.
	\end{equation}	
	
	\item Directed Leicht-Holme-Newman Index (DLHN):  Leicht-Holme-Newman Index scales the number of common neighbors by the product of the degrees of the two nodes. In the direct case, we scale the DCNE score of the couple $(i,j)$ by the product of the out-degree of $i$ and in-degree of $j$:
	\begin{equation} 
	\label{eq:dlhm}
	s^{DLHN}_{ij} = \dfrac {|\Gamma_{i}^{out} \cap \Gamma_{j}^{in}|}{\kappa_{i}^{out}  \kappa_{j}^{in}}.
	\end{equation}
	
	\item Directed Preferential Attachment Index (DPAT): The Preferential Attachment Index assigns a higher score to pairs of nodes if one or both have a high degree. In the directed network case, we focus on the out-degree of the first node and the in-degree node of the second:	
	\begin{equation} 
	\label{eq:dpat}
	s^{DPAT}_{ij} = {\kappa_{i}^{out}  \kappa_{j}^{in}}.
	\end{equation}
	
	\item  Directed Salton Index (DSAI): Salton Index mimics the cosine similarity between the two nodes. In the directed network case, we focus on the in and out neighbors of the two nodes:
	\begin{equation} 
	\label{eq:dsai}
	s^{DSAI}_{ij} = \frac {|\Gamma_{i}^{out} \cap \Gamma_{j}^{in}|}{\sqrt{\kappa_{i}^{out} \kappa_{j}^{in}}}.
	\end{equation}
	
	\item Directed Sorensen Index (DSOI):  Sorensen Index scales the Common Neighbors score by the sum of the degrees of the two nodes, which in the case of a directed network translates to scaling by the sum of the out-degree of the first node and the in-degree of the second:	
	\begin{equation} 
	\label{eq:dsoi}
	s^{DSOI}_{ij} = \frac {|\Gamma_{i}^{out} \cap \Gamma_{j}^{in}|}{\kappa_{i}^{out} + \kappa_{j}^{in}}.
	\end{equation}
	
\end{itemize}

\subsection{Experimental results}
Starting from a ground truth network, we randomly remove 10\% of existing links to use as a test set and present the remaining 90\%  to the link prediction algorithm. The latter uses the observed links to assign a score to non-observed edges to discriminate between removed links (false negative links) and originally non-existing links (true negative links). We repeat this process 1000 times for small networks having less than 1000 nodes and 100 times for large networks with more than 1000 nodes and report the performance measures averaged over all trials.

For small networks, we report all three performance measures: TPR, AUPR, and AUROC, whereas, for large networks, we report only TPR due to computational considerations.

\subsubsection{Effect of the horizon depth limit}
\label{sec:horizon}
\begin{figure*}[!t]
	\centering
	\includegraphics[width=0.25\textwidth]{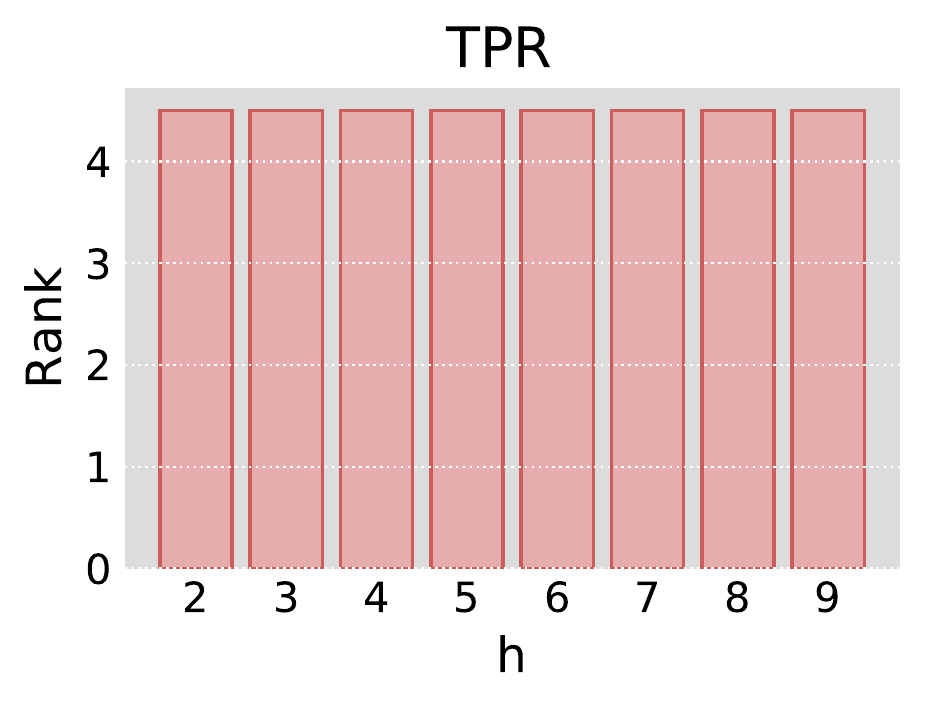}
	\includegraphics[width=0.25\textwidth]{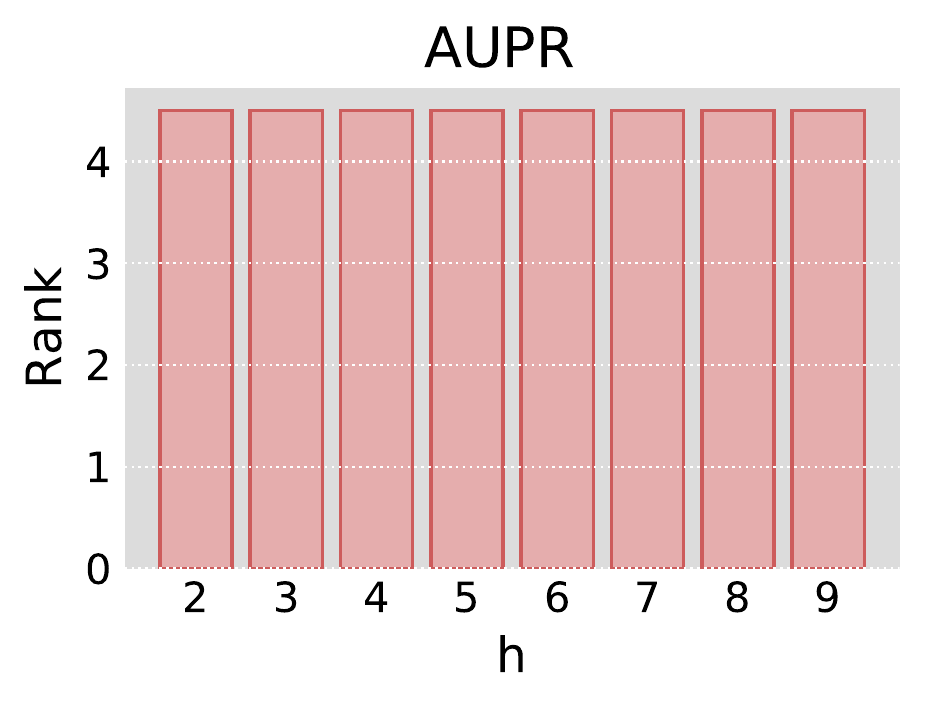}
	\includegraphics[width=0.25\textwidth]{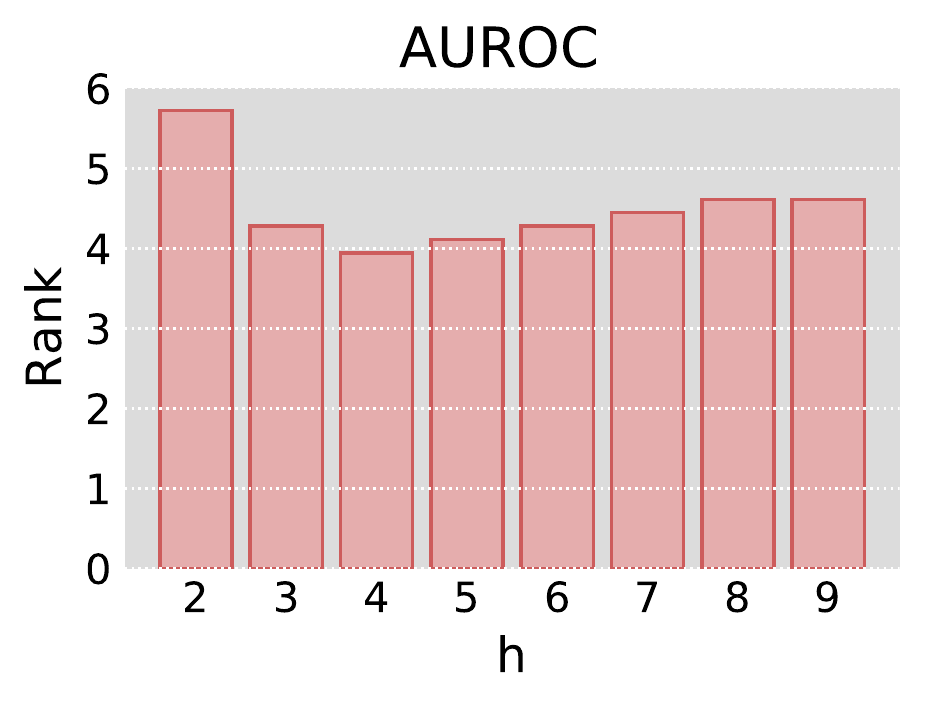}
	\caption{Effect of the horizon depth limit $h$ on the performance of Algorithm \ref{alg:1}.}
	\label{fig:horizon-alg1}
\end{figure*}

\begin{figure*}[!t]
	\centering
	\includegraphics[width=0.25\textwidth]{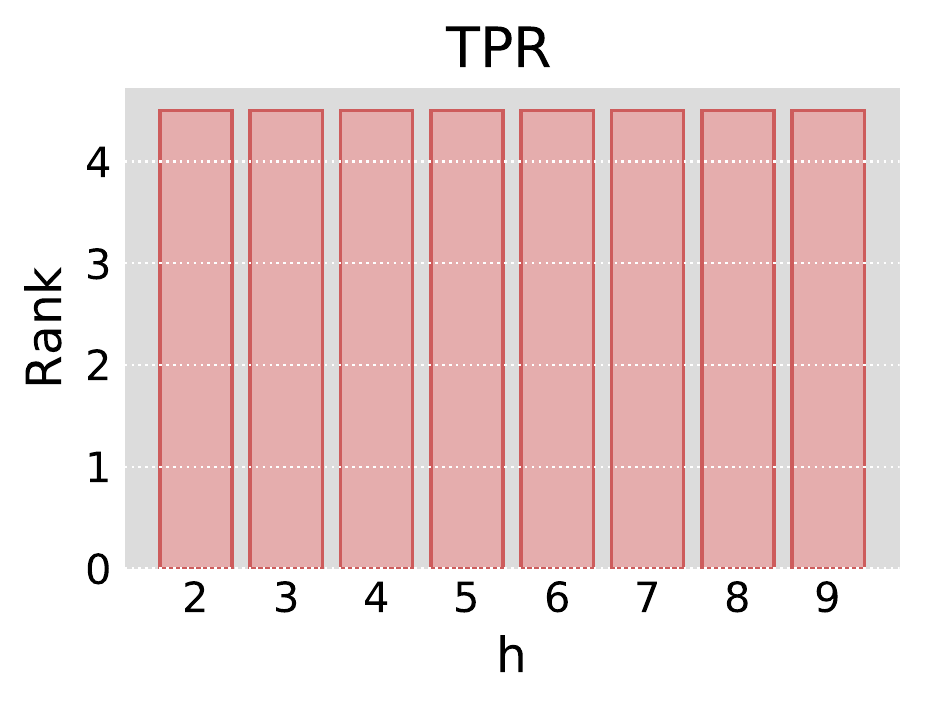}
	\includegraphics[width=0.25\textwidth]{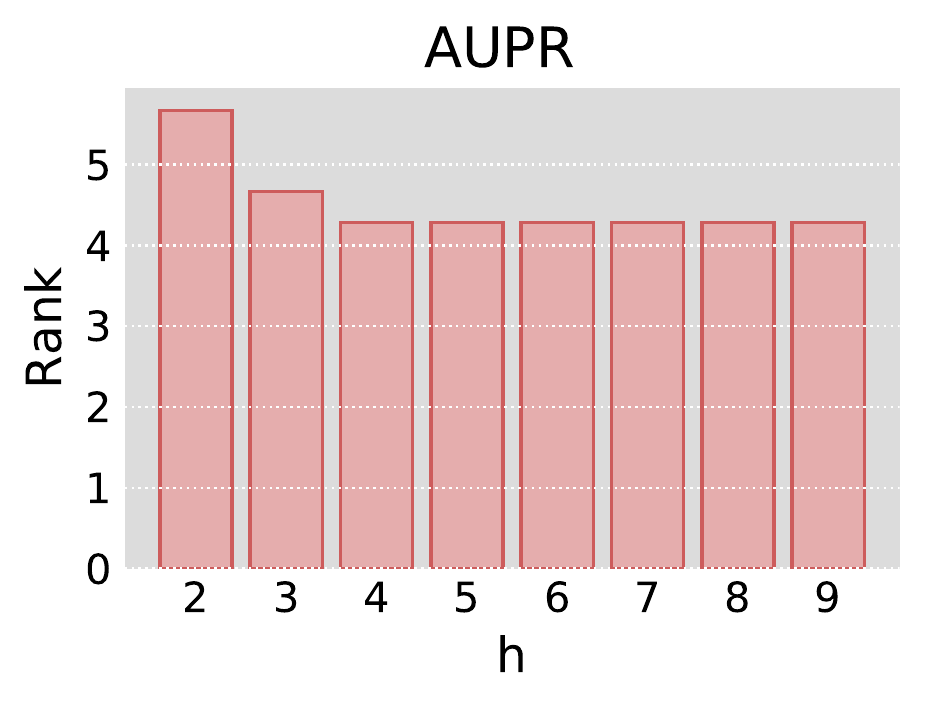}
	\includegraphics[width=0.25\textwidth]{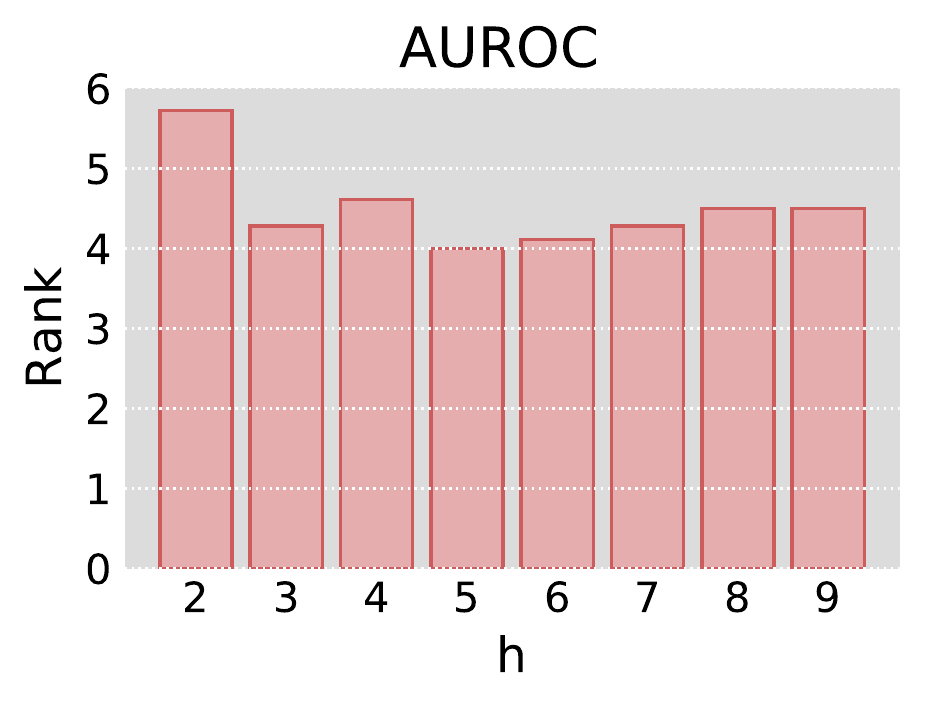}
	\caption{Effect of the horizon depth limit $h$ on the performance of Algorithm \ref{alg:2}.}
	\label{fig:horizon-alg2}
\end{figure*}

In this experiment, we measure the effect of the horizon depth limit on the performance of the algorithms. For this, we use nine small networks and vary the horizon limit $h$ from 2 to 9. Since most complex networks have a small radius due to the small world property, going beyond depth 9 will have no noticeable effect on the results. Figure \ref{fig:horizon-alg1} shows the average rank over all networks using TPR, AUPR and AUROC for Algorithm \ref{alg:1}, whereas Figure\ref{fig:horizon-alg2} shows the corresponding results for Algorithm \ref{alg:2}. 

We can see that the horizon depth has a noticeable effect on AUROC in both algorithms, with the best depth being 5. The AUPR does improve with larger horizon depths in Algorithm \ref{alg:2} with 4 being the best depth. On the other hand, TPR does not improve with deeper horizons in both algorithms. Since we are adopting TPR as our primary performance measure, we will set the horizon depth limit to 2 in the remaining experiments, a value that has the additional advantage of reducing the algorithm's computational complexity in terms of space and time.  


\subsubsection{Comparison against other algorithms}
In this experiment, we compare the proposed algorithms with the nine similarity-based link prediction algorithms presented in Section \ref{sec:directed}. We set the horizon depth limit to 2 as suggested in Section \ref{sec:horizon} and run all algorithms on two sets of networks: small networks and large networks. We use nine small networks having less than 1000 nodes on which we repeat the experiment 1000 times and report TPR, AUPR, and AUROC. We also use 14 large networks having more than 1000 nodes and up to 13,000 nodes. For each of these networks, we report TPR averaged over 100 trials.

Tables \ref{tab:tpr-small}, \ref{tab:aupr-small} and \ref{tab:auroc-small}  respectively show the TPR, AUPR, and AUROC obtained on small networks. Table \ref{tab:tpr-large} shows the TPR results obtained on large networks. The results show that Algorithm 2 achieves the best overall rank in small networks using all three performance measures, particularly TPR. It produces significant performance improvement compared to the other algorithms in most networks. Algorithm 2 is followed by directed Adamic-Adard Index and then Algorithm 1. Compared to small networks, the superiority of Algorithm 2 is more pronounced in large networks, where it gives the best results in 12 networks out of 14. In large networks, Algorithm \ref{alg:2} is followed in second position by directed Adamic-Adard Index, directed Common Neighbors, and then Algorithm 1.

To further assess the statistical significance of the obtained results,  we perform a two-sample, one-tailed Mann-Whitney-Wilcoxon test of the average significant ranks of TPR obtained on large networks. We run the test for each pair of algorithms and report the results in Table \ref{tab:tpr-wilcox}. The upper triangle of the table contains the test $ p $-values after adjustment using the Benjamini, Hochberg, and Yekutieli method. Results that are significant at 95\% confidence level are shown in bold. The lower triangular part reports the comparison of the average significant rank between the algorithm in the row and the one in the column if significant at $ p $-value =0.05. The results show that the superiority of Algorithm \ref{eq:alg2} over all other algorithms is statistically significant at a 95\% confidence level. 
\begin{table*}[!t]
	\caption{TPR results on small networks. The best performance with $p$-value 0.05 is indicated in bold. The last row shows the average significant rank of each algorithm; the lower, the better.}
	\label{tab:tpr-small}
	\centering
\begin{tabular}{lccccccccccc}
	\toprule
	Network                  &      DADA      & DCNE  & DHDI  & DHPI  & DJID  &  DLHN  &      DPAT      & DSAI  & DSOI  &      ALG1      &      ALG2      \\ \midrule
	Centrality Literature    &     0.159      & 0.158 & 0.028 & 0.027 & 0.026 & 0.007  & \textbf{0.186} & 0.030 & 0.026 &     0.157      &     0.142      \\ \midrule
	Chesapeake Middle        &     0.179      & 0.175 & 0.039 & 0.087 & 0.038 & 0.003  & \textbf{0.183} & 0.028 & 0.038 & \textbf{0.183} &     0.144      \\ \midrule
	Codeminer                &     0.055      & 0.031 & 0.005 & 0.028 & 0.005 & 0.002  &     0.004      & 0.004 & 0.005 &     0.063      & \textbf{0.068} \\ \midrule
	DNA Citation             & \textbf{0.016} & 0.004 & 0.001 & 0.008 & 0.000 & 0.001  &     0.006      & 0.000 & 0.000 & \textbf{0.018} & \textbf{0.017} \\ \midrule
	FilmTrust                &     0.164      & 0.146 & 0.013 & 0.044 & 0.013 & 0.007  &     0.052      & 0.013 & 0.010 &     0.064      & \textbf{0.260} \\ \midrule
	Japan Air                &     0.190      & 0.175 & 0.000 & 0.023 & 0.000 & 0.000  &     0.263      & 0.000 & 0.000 &     0.194      & \textbf{0.545} \\ \midrule
	Manufacturing e-mail     &     0.385      & 0.381 & 0.256 & 0.018 & 0.319 & 0.002  &     0.339      & 0.346 & 0.319 &     0.329      & \textbf{0.600} \\ \midrule
	Residence Hall           &     0.210      & 0.194 & 0.214 & 0.190 & 0.230 & 0.120  &     0.037      & 0.230 & 0.230 &     0.068      & \textbf{0.309} \\ \midrule
	US Air 97                &     0.215      & 0.193 & 0.014 & 0.013 & 0.004 & 0.001  &     0.208      & 0.003 & 0.004 & \textbf{0.272} &     0.243      \\ \midrule
	Average significant rank &     3.056      & 4.444 & 7.722 & 6.667 & 7.667 & 10.333 &     4.778      & 7.667 & 7.889 &     3.667      & \textbf{2.111} \\ \bottomrule
\end{tabular}
\end{table*}

\begin{table*}[!t]
	\caption{AUPR results on small networks. The best performance with $p$-value 0.05 is indicated in bold. The last row shows the average significant rank of each algorithm; the lower, the better.}
	\label{tab:aupr-small}
	\centering
\begin{tabular}{lccccccccccc}
	\toprule
	Network                  &      DADA      & DCNE  & DHDI  & DHPI  & DJID  &  DLHN  &      DPAT      & DSAI  & DSOI  &      ALG1      &      ALG2      \\ \midrule
	Centrality Literature    &     0.099      & 0.095 & 0.030 & 0.031 & 0.030 & 0.023  & \textbf{0.106} & 0.029 & 0.030 &     0.090      &     0.081      \\ \midrule
	Chesapeake Middle        &     0.099      & 0.100 & 0.040 & 0.077 & 0.044 & 0.032  & \textbf{0.117} & 0.046 & 0.043 &     0.109      &     0.094      \\ \midrule
	Codeminer                & \textbf{0.009} & 0.006 & 0.001 & 0.005 & 0.001 & 0.001  &     0.002      & 0.001 & 0.001 & \textbf{0.009} & \textbf{0.009} \\ \midrule
	DNA Citation             & \textbf{0.011} & 0.007 & 0.006 & 0.008 & 0.007 & 0.007  &     0.008      & 0.007 & 0.007 & \textbf{0.011} & \textbf{0.011} \\ \midrule
	FilmTrust                &     0.074      & 0.062 & 0.009 & 0.007 & 0.009 & 0.004  &     0.014      & 0.008 & 0.006 &     0.021      & \textbf{0.152} \\ \midrule
	Japan Air                &     0.125      & 0.114 & 0.012 & 0.020 & 0.012 & 0.010  &     0.232      & 0.012 & 0.012 &     0.158      & \textbf{0.523} \\ \midrule
	Manufacturing e-mail     &     0.371      & 0.366 & 0.199 & 0.066 & 0.260 & 0.030  &     0.335      & 0.288 & 0.260 &     0.292      & \textbf{0.573} \\ \midrule
	Residence Hall           &     0.143      & 0.129 & 0.129 & 0.098 & 0.143 & 0.064  &     0.015      & 0.145 & 0.143 &     0.038      & \textbf{0.235} \\ \midrule
	US Air 97                &     0.165      & 0.133 & 0.014 & 0.009 & 0.013 & 0.007  &     0.142      & 0.012 & 0.013 & \textbf{0.197} &     0.195      \\ \midrule
	Average significant rank &     2.722      & 4.444 & 8.389 & 7.167 & 7.722 & 10.333 &     4.278      & 7.000 & 8.056 &     3.667      & \textbf{2.222} \\ \bottomrule
\end{tabular}
\end{table*}

\begin{table*}[!t]
	\caption{AUROC results on small networks. The best performance with $p$-value 0.05 is indicated in bold. The last row shows the average significant rank of each algorithm; the lower, the better.}
	\label{tab:auroc-small}
	\centering
\begin{tabular}{lccccccccccc}
	\toprule
	Network                     &      DADA      &      DCNE      &      DHDI      &      DHPI      &      DJID      &      DLHN      &      DPAT      &      DSAI      &      DSOI      &      ALG1      &      ALG2      \\ \midrule
	Centrality Literature       &     0.766      &     0.765      &     0.759      &     0.758      &     0.758      &     0.754      & \textbf{0.859} &     0.758      &     0.758      &     0.768      &     0.765      \\ \midrule
	Chesapeake Middle           &     0.763      &     0.755      &     0.700      &     0.748      &     0.710      &     0.689      & \textbf{0.807} &     0.721      &     0.710      &     0.772      &     0.760      \\ \midrule
	Codeminer                   & \textbf{0.543} & \textbf{0.543} & \textbf{0.543} & \textbf{0.543} & \textbf{0.543} & \textbf{0.543} & \textbf{0.544} & \textbf{0.543} & \textbf{0.543} & \textbf{0.543} & \textbf{0.543} \\ \midrule
	DNA Citation                & \textbf{0.548} & \textbf{0.547} & \textbf{0.545} & \textbf{0.547} & \textbf{0.546} & \textbf{0.546} &     0.530      & \textbf{0.546} & \textbf{0.546} & \textbf{0.548} & \textbf{0.548} \\ \midrule
	FilmTrust                   &     0.709      &     0.709      &     0.708      &     0.707      &     0.708      &     0.707      &     0.683      &     0.708      &     0.708      &     0.709      & \textbf{0.709} \\ \midrule
	Japan Air                   &     0.773      &     0.763      &     0.505      &     0.646      &     0.510      &     0.399      &     0.871      &     0.525      &     0.510      &     0.853      & \textbf{0.895} \\ \midrule
	Manufacturing e-mail        &     0.925      &     0.923      &     0.874      &     0.817      &     0.896      &     0.627      &     0.912      &     0.910      &     0.896      &     0.920      & \textbf{0.964} \\ \midrule
	Residence Hall              &     0.893      &     0.887      &     0.887      &     0.889      &     0.891      &     0.876      &     0.648      &     0.892      &     0.891      &     0.858      & \textbf{0.909} \\ \midrule
	US Air 97                   &     0.854      &     0.848      &     0.830      &     0.816      &     0.829      &     0.801      & \textbf{0.874} &     0.825      &     0.829      &     0.859      &     0.856      \\ \midrule
	Average significant ranking &     3.667      &     4.944      &     7.556      &     7.389      &     6.889      &     9.500      &     5.444      &     6.611      &     6.889      &     4.278      & \textbf{2.833} \\ \bottomrule
\end{tabular}
\end{table*}

\begin{table*}[!t]
	\caption{TPR results on large networks. The best performance with $p$-value 0.05 is indicated in bold. The last row shows the average significant rank of each algorithm; the lower, the better.}
	\label{tab:tpr-large}
	\centering
\begin{tabular}{lccccccccccc}
	\toprule
	Network                  &      DADA      & DCNE  & DHDI  & DHPI  & DJID  & DLHN  & DPAT  & DSAI  & DSOI  & ALG1  &      ALG2      \\ \midrule
	Adolescent Health        &     0.078      & 0.075 & 0.057 & 0.020 & 0.059 & 0.025 & 0.001 & 0.056 & 0.059 & 0.013 & \textbf{0.163} \\ \midrule
	Advogato                 &     0.129      & 0.119 & 0.020 & 0.013 & 0.009 & 0.003 & 0.044 & 0.007 & 0.009 & 0.055 & \textbf{0.171} \\ \midrule
	BitcoinAlpha             &     0.088      & 0.084 & 0.000 & 0.004 & 0.000 & 0.000 & 0.051 & 0.000 & 0.000 & 0.051 & \textbf{0.273} \\ \midrule
	Ciao                     &     0.115      & 0.114 & 0.052 & 0.002 & 0.042 & 0.000 & 0.066 & 0.023 & 0.042 & 0.051 & \textbf{0.197} \\ \midrule
	Criminal                 &     0.041      & 0.041 & 0.000 & 0.004 & 0.000 & 0.000 & 0.021 & 0.000 & 0.000 & 0.036 & \textbf{0.093} \\ \midrule
	DNC Email                &     0.206      & 0.199 & 0.000 & 0.006 & 0.000 & 0.000 & 0.074 & 0.000 & 0.000 & 0.135 & \textbf{0.226} \\ \midrule
	FOLDOC                   &     0.234      & 0.210 & 0.252 & 0.240 & 0.268 & 0.240 & 0.010 & 0.271 & 0.268 & 0.016 & \textbf{0.396} \\ \midrule
	Human Protein            &     0.001      & 0.001 & 0.000 & 0.000 & 0.000 & 0.000 & 0.011 & 0.000 & 0.000 & 0.003 & \textbf{0.026} \\ \midrule
	Indochina 2004           & \textbf{0.539} & 0.471 & 0.145 & 0.111 & 0.100 & 0.002 & 0.009 & 0.088 & 0.100 & 0.332 &     0.515      \\ \midrule
	Mathoverflow C2A         & \textbf{0.106} & 0.103 & 0.000 & 0.002 & 0.000 & 0.000 & 0.080 & 0.000 & 0.000 & 0.054 &     0.092      \\ \midrule
	ODLIS                    &     0.104      & 0.087 & 0.014 & 0.020 & 0.015 & 0.007 & 0.036 & 0.013 & 0.015 & 0.069 & \textbf{0.127} \\ \midrule
	Web EPA                  &     0.028      & 0.016 & 0.001 & 0.007 & 0.001 & 0.000 & 0.012 & 0.001 & 0.001 & 0.028 & \textbf{0.031} \\ \midrule
	Web Edu                  &     0.199      & 0.161 & 0.041 & 0.048 & 0.027 & 0.000 & 0.007 & 0.024 & 0.027 & 0.333 & \textbf{0.398} \\ \midrule
	WikiTalk                 &     0.101      & 0.094 & 0.000 & 0.006 & 0.000 & 0.000 & 0.060 & 0.000 & 0.000 & 0.096 & \textbf{0.127} \\ \midrule
	Average significant rank &     2.571      & 3.643 & 6.821 & 6.536 & 7.500 & 9.893 & 6.179 & 8.393 & 8.500 & 4.750 & \textbf{1.214} \\ \bottomrule
\end{tabular}
\end{table*}

\begin{table*}[!t]
	\centering
	\caption{Results of the Mann-Whitney-Wilcoxon test of the mean significant rank based on TPR in large networks.}
	\label{tab:tpr-wilcox}
\begin{tabular}{lccccccccccc}
	\toprule
	     & DADA &     DCNE      &     DHDI      &     DHPI      &     DJID      &     DLHN      &     DPAT      &     DSAI      &     DSOI      &     ALG1      &     ALG2      \\ \midrule
	DADA &      & \textbf{0.03} & \textbf{0.00} & \textbf{0.00} & \textbf{0.00} & \textbf{0.00} & \textbf{0.00} & \textbf{0.00} & \textbf{0.00} & \textbf{0.01} & \textbf{0.01} \\ \midrule
	DCNE & $>$  &               & \textbf{0.00} & \textbf{0.00} & \textbf{0.00} & \textbf{0.00} & \textbf{0.01} & \textbf{0.00} & \textbf{0.00} &     0.37      & \textbf{0.00} \\ \midrule
	DHDI & $>$  &      $>$      &               &     1.00      &     0.53      & \textbf{0.00} &     0.63      & \textbf{0.02} &     0.13      & \textbf{0.02} & \textbf{0.00} \\ \midrule
	DHPI & $>$  &      $>$      &               &               &     0.17      & \textbf{0.00} &     0.35      & \textbf{0.02} &     0.05      & \textbf{0.02} & \textbf{0.00} \\ \midrule
	DJID & $>$  &      $>$      &               &               &               & \textbf{0.00} &     1.00      &     0.13      &     0.82      & \textbf{0.04} & \textbf{0.00} \\ \midrule
	DLHN & $>$  &      $>$      &      $>$      &      $>$      &      $>$      &               & \textbf{0.02} &     0.08      &     0.43      & \textbf{0.00} & \textbf{0.00} \\ \midrule
	DPAT & $>$  &      $>$      &               &               &               &      $<$      &               &     0.48      &     0.38      &     0.30      & \textbf{0.00} \\ \midrule
	DSAI & $>$  &      $>$      &      $>$      &      $>$      &               &               &               &               &     1.00      & \textbf{0.03} & \textbf{0.00} \\ \midrule
	DSOI & $>$  &      $>$      &               &               &               &               &               &               &               & \textbf{0.01} & \textbf{0.00} \\ \midrule
	ALG1 & $>$  &               &      $<$      &      $<$      &      $<$      &      $<$      &               &      $<$      &      $<$      &               & \textbf{0.00} \\ \midrule
	ALG2 & $<$  &      $<$      &      $<$      &      $<$      &      $<$      &      $<$      &      $<$      &      $<$      &      $<$      &      $<$      &               \\ \bottomrule
\end{tabular}
\end{table*}

\section{Conclusion}
\label{sec:conclusion}
Link prediction in complex networks aims to predict the future links that can form between nodes based on the currently observed connections. Link prediction is a problem with significant theoretical and practical value and has received increasing attention from the research community.  
The drive behind this interest is the need to analyze and predict the evolution of networked systems and deal with incomplete observations caused by the shortcomings typically encountered in the data collection process.

Due to the complexity and difficulty of the link prediction task, many prediction approaches have been proposed by researchers, particularly for undirected complex networks. However, not all real systems can be represented as undirected networks, and treating them as such inevitably results in information loss and, consequently, the degradation of the algorithms' predictive power. Despite the widespread of directed networks in real-life data, research effort in developing link prediction algorithms tailored to directed systems remains limited.

This paper introduced a novel approach to tackling the link prediction problem in directed networks. The proposed approach builds on the recent success of similarity-popularity methods in undirected networks. The presented algorithms are designed to handle and harness the asymmetry in node relationships by modeling it as asymmetry in similarity and popularity. Given the observed network topology, the algorithms estimate hidden similarities using shortest path distances based on an edge weight map designed to capture and factor out the asymmetry in node relations. 
An extensive experimental evaluation of the proposed approach demonstrates its effectiveness in predicting missing links in real-life networks of various types and sizes.  

In future work, we propose to develop a network embedding framework tailored for directed networks based on the presented similarity-popularity link prediction method. Such embedding can be used for detecting spurious links and other network analysis tasks such as community and pattern detection.


\ifCLASSOPTIONcompsoc
  \section*{Acknowledgments}
      This research work is supported by the Research Center, CCIS, King Saud University, Riyadh, Saudi Arabia.
\else
  \section*{Acknowledgment}
\fi

\ifCLASSOPTIONcaptionsoff
  \newpage
\fi



\bibliographystyle{IEEEtran}

\begin{thebibliography}{10}
	\providecommand{\url}[1]{#1}
	\csname url@samestyle\endcsname
	\providecommand{\newblock}{\relax}
	\providecommand{\bibinfo}[2]{#2}
	\providecommand{\BIBentrySTDinterwordspacing}{\spaceskip=0pt\relax}
	\providecommand{\BIBentryALTinterwordstretchfactor}{4}
	\providecommand{\BIBentryALTinterwordspacing}{\spaceskip=\fontdimen2\font plus
		\BIBentryALTinterwordstretchfactor\fontdimen3\font minus
		\fontdimen4\font\relax}
	\providecommand{\BIBforeignlanguage}[2]{{%
			\expandafter\ifx\csname l@#1\endcsname\relax
			\typeout{** WARNING: IEEEtran.bst: No hyphenation pattern has been}%
			\typeout{** loaded for the language `#1'. Using the pattern for}%
			\typeout{** the default language instead.}%
			\else
			\language=\csname l@#1\endcsname
			\fi
			#2}}
	\providecommand{\BIBdecl}{\relax}
	\BIBdecl
	
	\bibitem{albert2002statistical1}
	R.~Albert and A.-L. Barab{\'a}si, ``Statistical mechanics of complex
	networks,'' \emph{Reviews of modern physics}, vol.~74, no.~1, p.~47, 2002.
	
	\bibitem{guimera2009missing}
	R.~Guimer{\`a} and M.~Sales-Pardo, ``Missing and spurious interactions and the
	reconstruction of complex networks,'' \emph{Proceedings of the National
		Academy of Sciences}, vol. 106, no.~52, pp. 22\,073--22\,078, 2009.
	
	\bibitem{lu2011link}
	L.~L{\"u} and T.~Zhou, ``Link prediction in complex networks: A survey,''
	\emph{Physica A: statistical mechanics and its applications}, vol. 390,
	no.~6, pp. 1150--1170, 2011.
	
	\bibitem{Boguna2021}
	M.~Bogu{\~{n}}{\'a}, I.~Bonamassa, M.~De~Domenico, S.~Havlin, D.~Krioukov, and
	M.~{\'A}. Serrano, ``Network geometry,'' \emph{Nature Reviews Physics},
	vol.~3, no.~2, pp. 114--135, Feb 2021.
	
	\bibitem{kerrache20}
	S.~Kerrache, R.~Alharbi, and H.~Benhidour, ``\BIBforeignlanguage{eng}{A
		scalable similarity-popularity link prediction method},''
	\emph{\BIBforeignlanguage{eng}{Scientific reports}}, vol.~10, no.~1, pp.
	6394--6394, 2020.
	
	\bibitem{huang2005link}
	Z.~Huang, X.~Li, and H.~Chen, ``Link prediction approach to collaborative
	filtering,'' in \emph{Proceedings of the 5th ACM/IEEE-CS joint conference on
		Digital libraries}.\hskip 1em plus 0.5em minus 0.4em\relax ACM, 2005, pp.
	141--142.
	
	\bibitem{li2009recommendation}
	X.~Li and H.~Chen, ``Recommendation as link prediction: a graph kernel-based
	machine learning approach,'' in \emph{Proceedings of the 9th ACM/IEEE-CS
		joint conference on Digital libraries}.\hskip 1em plus 0.5em minus
	0.4em\relax ACM, 2009, pp. 213--216.
	
	\bibitem{liu2007predicting}
	Y.~Liu and Z.~Kou, ``Predicting who rated what in large-scale datasets,''
	\emph{ACM SIGKDD Explorations Newsletter}, vol.~9, no.~2, pp. 62--65, 2007.
	
	\bibitem{adamic2003friends}
	L.~A. Adamic and E.~Adar, ``Friends and neighbors on the web,'' \emph{Social
		networks}, vol.~25, no.~3, pp. 211--230, 2003.
	
	\bibitem{newman2001clustering}
	M.~E.~J. Newman, ``Clustering and preferential attachment in growing
	networks,'' \emph{Phys. Rev. E}, vol.~64, Jul 2001.
	
	\bibitem{ravasz2002hierarchical}
	E.~Ravasz, A.~L. Somera, D.~A. Mongru, Z.~N. Oltvai, and A.-L. Barab{\'a}si,
	``Hierarchical organization of modularity in metabolic networks,''
	\emph{science}, vol. 297, no. 5586, pp. 1551--1555, 2002.
	
	\bibitem{jaccard1901etude}
	P.~Jaccard, ``{\'E}tude comparative de la distribution florale dans une portion
	des alpes et des jura,'' \emph{Bull Soc Vaudoise Sci Nat}, vol.~37, pp.
	547--579, 1901.
	
	\bibitem{dillon1983introduction}
	M.~Dillon, ``Introduction to modern information retrieval,'' 1983.
	
	\bibitem{sorensen1948method}
	T.~S{\o}rensen, ``A method of establishing groups of equal amplitude in plant
	sociology based on similarity of species and its application to analyses of
	the vegetation on danish commons,'' \emph{Biol. Skr.}, vol.~5, pp. 1--34,
	1948.
	
	\bibitem{zhou2009predicting}
	T.~Zhou, L.~L{\"u}, and Y.-C. Zhang, ``Predicting missing links via local
	information,'' \emph{The European Physical Journal B}, vol.~71, no.~4, pp.
	623--630, 2009.
	
	\bibitem{lu2009similarity}
	L.~L{\"u}, C.-H. Jin, and T.~Zhou, ``Similarity index based on local paths for
	link prediction of complex networks,'' \emph{Physical Review E}, vol.~80,
	no.~4, p. 046122, 2009.
	
	\bibitem{katz1953new}
	L.~Katz, ``A new status index derived from sociometric analysis,''
	\emph{Psychometrika}, vol.~18, no.~1, pp. 39--43, 1953.
	
	\bibitem{liu2010link}
	W.~Liu and L.~L{\"u}, ``Link prediction based on local random walk,'' \emph{EPL
		(Europhysics Letters)}, vol.~89, no.~5, p. 58007, 2010.
	
	\bibitem{tong2006fast}
	H.~Tong, C.~Faloutsos, and J.-Y. Pan, ``Fast random walk with restart and its
	applications,'' pp. 613--622, 2006.
	
	\bibitem{fouss07}
	F.~Fouss, A.~Pirotte, J.-m. Renders, and M.~Saerens, ``Random-walk computation
	of similarities between nodes of a graph with application to collaborative
	recommendation,'' \emph{IEEE Transactions on Knowledge and Data Engineering},
	vol.~19, no.~3, pp. 355--369, 2007.
	
	\bibitem{vanunu2008propagation}
	O.~Vanunu and R.~Sharan, ``A propagation-based algorithm for inferring
	gene-disease assocations.'' in \emph{German Conference on Bioinformatics},
	2008, pp. 54--52.
	
	\bibitem{jeh2002simrank}
	G.~Jeh and J.~Widom, ``Simrank: a measure of structural-context similarity,''
	in \emph{Proceedings of the eighth ACM SIGKDD international conference on
		Knowledge discovery and data mining}.\hskip 1em plus 0.5em minus 0.4em\relax
	ACM, 2002, pp. 538--543.
	
	\bibitem{goldenberg2010survey}
	A.~Goldenberg, A.~X. Zheng, S.~E. Fienberg, E.~M. Airoldi \emph{et~al.}, ``A
	survey of statistical network models,'' \emph{Foundations and
		Trends{\textregistered} in Machine Learning}, vol.~2, no.~2, pp. 129--233,
	2010.
	
	\bibitem{clauset2008hierarchical}
	A.~Clauset, C.~Moore, and M.~E. Newman, ``Hierarchical structure and the
	prediction of missing links in networks,'' \emph{Nature}, vol. 453, no. 7191,
	p.~98, 2008.
	
	\bibitem{liu2013correlations}
	Z.~Liu, J.-L. He, K.~Kapoor, and J.~Srivastava, ``Correlations between
	community structure and link formation in complex networks,'' \emph{PloS
		one}, vol.~8, no.~9, p. e72908, 2013.
	
	\bibitem{huang2010link}
	Z.~Huang, ``Link prediction based on graph topology: The predictive value of
	generalized clustering coefficient,'' 2010.
	
	\bibitem{al2006link}
	M.~Al~Hasan, V.~Chaoji, S.~Salem, and M.~Zaki, ``Link prediction using
	supervised learning,'' in \emph{SDM06: workshop on link analysis,
		counter-terrorism and security}, 2006.
	
	\bibitem{doppa2010learning}
	J.~R. Doppa, J.~Yu, P.~Tadepalli, and L.~Getoor, ``Learning algorithms for link
	prediction based on chance constraints,'' in \emph{Joint european conference
		on machine learning and knowledge discovery in databases}.\hskip 1em plus
	0.5em minus 0.4em\relax Springer, 2010, pp. 344--360.
	
	\bibitem{boguna2009navigability}
	M.~Boguna, D.~Krioukov, and K.~C. Claffy, ``Navigability of complex networks,''
	\emph{Nature Physics}, vol.~5, no.~1, p.~74, 2009.
	
	\bibitem{papadopoulos2012popularity}
	F.~Papadopoulos, M.~Kitsak, M.~{\'A}. Serrano, M.~Bogun{\'a}, and D.~Krioukov,
	``Popularity versus similarity in growing networks,'' \emph{Nature}, vol.
	489, no. 7417, p. 537, 2012.
	
	\bibitem{zou2014exploiting}
	J.~Zou and F.~Fekri, ``Exploiting popularity and similarity for link
	recommendation in twitter networks.'' in \emph{RSWeb@ RecSys}, 2014.
	
	\bibitem{fastlp}
	H.~B. Alharbi, Ruwayda and S.~Kerrache, ``Scalable link prediction in complex
	networks using a type of geodesic distance,'' in \emph{Asia Multi Conference
		on Mathematical Modelling and Computer Simulatio}, 2016.
	
	\bibitem{schall2014link}
	D.~Schall, ``Link prediction in directed social networks,'' \emph{Social
		Network Analysis and Mining}, vol.~4, no.~1, p. 157, 2014.
	
	\bibitem{garcia2014link}
	D.~Garcia~Gasulla and C.~U. Cort{\'e}s~Garc{\'\i}a, ``Link prediction in very
	large directed graphs: Exploiting hierarchical properties in parallel,'' in
	\emph{Proceedings of the 3rd Workshop on Knowledge Discovery and Data Mining
		Meets Linked Open Data co-located with 11th Extended Semantic Web Conference
		(ESWC 2014): Crete, Greece, May 25, 2014}.\hskip 1em plus 0.5em minus
	0.4em\relax CEUR-WS. org, 2014, pp. 1--13.
	
	\bibitem{Serrano2008Self-similarityofComplexNetworksandHiddenMetricSpaces}
	M.~A. Serrano, D.~Krioukov, and M.~Bogun{\'a}, ``Self-similarity of complex
	networks and hidden metric spaces,'' \emph{Physical review letters}, vol.
	100, no.~7, p. 078701, 2008.
	
	\bibitem{Ortiz2017}
	E.~Ortiz, M.~Starnini, and M.~{\'A}. Serrano, ``Navigability of temporal
	networks in hyperbolic space,'' \emph{Scientific Reports}, vol.~7, no.~1, p.
	15054, Nov 2017.
	
	\bibitem{snapnets}
	J.~Leskovec and A.~Krevl, ``{SNAP Datasets}: {Stanford} large network dataset
	collection,'' \url{http://snap.stanford.edu/data}, Jun. 2014.
	
	\bibitem{konect}
	J.~Kunegis, ``\BIBforeignlanguage{eng}{Konect: the koblenz network
		collection},'' in \emph{\BIBforeignlanguage{eng}{Proceedings of the 22nd
			International Conference on world wide web}}, ser. WWW '13 Companion.\hskip
	1em plus 0.5em minus 0.4em\relax ACM, 2013, pp. 1343--1350.
	
	\bibitem{pajaknets}
	V.~Batagelj and A.~Mrvar, ``Pajek datasets,''
	\url{http://vlado.fmf.uni-lj.si/pub/networks/data}, 2006.
	
	\bibitem{adolescent-health}
	J.~Moody, ``\BIBforeignlanguage{eng}{Peer influence groups: identifying dense
		clusters in large networks},'' \emph{\BIBforeignlanguage{eng}{Social
			networks}}, vol.~23, no.~4, pp. 261--283, 2001.
	
	\bibitem{advogato}
	P.~Massa, M.~Salvetti, and D.~Tomasoni, ``Bowling alone and trust decline in
	social network sites,'' in \emph{Proc. Int. Conf. Dependable, Autonomic and
		Secure Computing}, 2009, pp. 658--663.
	
	\bibitem{bitcoinalpha}
	S.~Kumar, F.~Spezzano, V.~S. Subrahmanian, and C.~Faloutsos,
	``\BIBforeignlanguage{eng}{Edge weight prediction in weighted signed
		networks},'' in \emph{\BIBforeignlanguage{eng}{2016 IEEE 16th International
			Conference on Data Mining (ICDM)}}.\hskip 1em plus 0.5em minus 0.4em\relax
	IEEE, 2016, pp. 221--230.
	
	\bibitem{centrality-90}
	N.~Hummon, P.~Doreian, and L.~Freeman, ``Analyzing the structure of the
	centrality-productivity literature created between 1948 and 1979,''
	\emph{Knowledge}, vol.~11, pp. 459--480, 06 1990.
	
	\bibitem{chesapeake2002}
	J.~D. Hagy, ``Eutrophication, hypoxia and trophic transfer efficiency in
	chesapeake bay,'' Ph.D. dissertation, University of Maryland at College Park
	(USA), 2002.
	
	\bibitem{ciao}
	G.~Guo, J.~Zhang, D.~Thalmann, and N.~Yorke-Smith,
	``\BIBforeignlanguage{eng}{Etaf: An extended trust antecedents framework for
		trust prediction},'' in \emph{\BIBforeignlanguage{eng}{2014 IEEE/ACM
			International Conference on Advances in Social Networks Analysis and Mining
			(ASONAM 2014)}}.\hskip 1em plus 0.5em minus 0.4em\relax IEEE, 2014, pp.
	540--547.
	
	\bibitem{Heymannm2008Javacode}
	S.~S.Heymannm and J.~Palmier, ``Source code structure of a java program,''
	2008.
	
	\bibitem{criminal}
	E.~Mainas, ``The analysis of criminal and terrorist organisations as social
	network structures,'' Master's thesis, Institute of Criminal Justice Studies,
	University of Portsmouth, UK, 2009.
	
	\bibitem{Hummon1989ConnectivityInCitationNetworkDNA}
	N.~P. Hummon and P.~Dereian, ``Connectivity in a citation network: The
	development of dna theory,'' \emph{Social Networks}, vol.~11, no.~1, pp.
	39--63, 1989.
	
	\bibitem{filmtrust}
	G.~Guo, J.~Zhang, and N.~Yorke-Smith, ``A novel evidence-based bayesian
	similarity measure for recommender systems,'' \emph{ACM Trans. Web}, vol.~10,
	no.~2, may 2016.
	
	\bibitem{foldoc}
	B.~Vladimir, M.~Andrej, and Z.~Matjaz, ``Network analysis of dictionaries,'' in
	\emph{Language Technologies, Ljubljana}, 2002, pp. 135--142.
	
	\bibitem{human-protein}
	U.~Stelzl, U.~Worm, M.~Lalowski, C.~Haenig, F.~H. Brembeck, H.~Goehler,
	M.~Stroedicke, M.~Zenkner, A.~Schoenherr, S.~Koeppen, J.~Timm, S.~Mintzlaff,
	C.~Abraham, N.~Bock, S.~Kietzmann, A.~Goedde, E.~Toksöz, A.~Droege,
	S.~Krobitsch, B.~Korn, W.~Birchmeier, H.~Lehrach, and E.~E. Wanker,
	``\BIBforeignlanguage{eng}{A human protein-protein interaction network: A
		resource for annotating the proteome},''
	\emph{\BIBforeignlanguage{eng}{Cell}}, vol. 122, no.~6, pp. 957--968, 2005.
	
	\bibitem{bovwfi}
	P.~Boldi and S.~Vigna, ``The {W}eb{G}raph framework {I}: {C}ompression
	techniques,'' in \emph{Proc. of the Thirteenth International World Wide Web
		Conference (WWW 2004)}.\hskip 1em plus 0.5em minus 0.4em\relax Manhattan,
	USA: ACM Press, 2004, pp. 595--601.
	
	\bibitem{brsllp}
	P.~Boldi, M.~Rosa, M.~Santini, and S.~Vigna, ``Layered label propagation: A
	multiresolution coordinate-free ordering for compressing social networks,''
	in \emph{Proceedings of the 20th international conference on World Wide Web},
	S.~Srinivasan, K.~Ramamritham, A.~Kumar, M.~P. Ravindra, E.~Bertino, and
	R.~Kumar, Eds.\hskip 1em plus 0.5em minus 0.4em\relax ACM Press, 2011, pp.
	587--596.
	
	\bibitem{guimera2005}
	R.~Guimer{\`{a}}, S.~Mossa, A.~Turtschi, and L.~A.~N. Amaral, ``{The worldwide
		air transportation network: Anomalous centrality, community structure, and
		cities' global roles.}'' \emph{Proceedings of the National Academy of
		Sciences of the United States of America}, vol. 102, no.~22, pp. 7794--9, may
	2005.
	
	\bibitem{michalski2011matching}
	R.~Michalski, S.~Palus, and P.~Kazienko, ``Matching organizational structure
	and social network extracted from email communication,'' in \emph{Business
		Information Systems}.\hskip 1em plus 0.5em minus 0.4em\relax Springer, 2011,
	pp. 197--206.
	
	\bibitem{paranjape17}
	A.~Paranjape, A.~R. Benson, and J.~Leskovec, ``Motifs in temporal networks,''
	in \emph{Proceedings of the Tenth ACM International Conference on Web Search
		and Data Mining}, ser. WSDM '17.\hskip 1em plus 0.5em minus 0.4em\relax New
	York, NY, USA: ACM, 2017, pp. 601--610.
	
	\bibitem{Reitz2002}
	J.~M. Reitz, ``Odlis: Online dictionary of library and information science,''
	2002.
	
	\bibitem{residencehall98}
	L.~Freeman, C.~Webster, and D.~Kirke, ``\BIBforeignlanguage{English}{Exploring
		social structure using dynamic three-dimensional color images},''
	\emph{\BIBforeignlanguage{English}{Social Networks}}, vol.~20, no.~2, pp.
	109--118, 4 1998.
	
	\bibitem{gleich04}
	D.~Gleich, L.~Zhukov, and P.~Berkhin, ``Fast parallel pagerank: A linear system
	approach,'' \emph{Yahoo! Research Technical Report YRL-2004-038}, vol.~13,
	p.~22, 2004.
	
	\bibitem{pajek11}
	W.~De~Nooy, A.~Mrvar, and V.~Batagelj, \emph{Exploratory social network
		analysis with Pajek}.\hskip 1em plus 0.5em minus 0.4em\relax Cambridge
	University Press, 2011, vol.~27.
	
	\bibitem{wikitalk}
	J.~Sun, J.~Kunegis, and S.~Staab, ``\BIBforeignlanguage{eng}{Predicting user
		roles in social networks using transfer learning with feature
		transformation},'' in \emph{\BIBforeignlanguage{eng}{arXiv.org}}.\hskip 1em
	plus 0.5em minus 0.4em\relax Ithaca: Cornell University Library, arXiv.org,
	2016.
	
	\bibitem{Lu2011LinkPredictioninComplexNetworks}
	L.~L{\"u} and T.~Zhou, ``Link prediction in complex networks: A survey,''
	\emph{Physica A: Statistical Mechanics and its Applications}, vol. 390,
	no.~6, pp. 1150--1170, 2011.
	
	\bibitem{yang2015EvaluatingLinkPredictionMethods}
	Y.~Yang, R.~N. Lichtenwalter, and N.~V. Chawla, ``Evaluating link prediction
	methods,'' \emph{Knowledge and Information Systems}, vol.~45, no.~3, pp.
	751--782, 2015.
	
	\bibitem{garcia2016limitations}
	D.~Garcia-Gasulla, E.~Ayguad{\'e}, J.~Labarta, and U.~Cort{\'e}s, ``Limitations
	and alternatives for the evaluation of large-scale link prediction,''
	preprint at http://arxiv.org/abs/1611.00547 (2016).
	
	\bibitem{Wang2016}
	W.~Wang, F.~Cai, P.~Jiao, and L.~Pan, ``A perturbation-based framework for link
	prediction via non-negative matrix factorization,'' \emph{Scientific
		reports}, vol.~6, 2016.
	
	\bibitem{muscoloni2017LocalRing}
	A.~Muscoloni and C.~V. Cannistraci, ``Local-ring network automata and the
	impact of hyperbolic geometry in complex network link-prediction,'' 2017,
	preprint at http://arXiv:1707.09496 [physics.soc-ph].
	
	\bibitem{Davis2006TheRelationshipBetweenPrecision-RecallandROCCurves}
	J.~Davis and M.~Goadrich, ``The relationship between precision-recall and roc
	curves,'' in \emph{Proceedings of the 23rd international conference on
		Machine learning}.\hskip 1em plus 0.5em minus 0.4em\relax ACM, 2006, pp.
	233--240.
	
\end{thebibliography}
%

%




\end{document}